\documentclass[pre,showpacs,twocolumn,superscriptaddress]{revtex4}
\usepackage{amssymb,graphicx,amsbsy}
\usepackage{threeparttable}

\begin{document}

\title{Surface and bulk criticality in midpoint percolation}
\author{Seung Ki Baek}
\email[Corresponding author, E-mail: ]{garuda@tp.umu.se}
\affiliation{Department of Physics, Ume{\aa} University, 901 87 Ume{\aa},
Sweden}
\author{Petter Minnhagen}
\affiliation{Department of Physics, Ume{\aa} University, 901 87 Ume{\aa},
Sweden}
\author{Beom Jun Kim}
\affiliation{BK21 Physics Research Division and Department of Physics,
Sungkyunkwan University, Suwon 440-746, Korea}

\begin{abstract}
The concept of midpoint percolation has recently been applied to
characterize the double percolation transitions in negatively curved
structures. Regular $d$-dimensional hypercubic lattices are in the present
work investigated using the same concept.
Specifically, the site-percolation transitions at the critical thresholds are
investigated for dimensions up to $d=10$ by means of the Leath algorithm.
It is shown that the explicit inclusion of the boundaries 
provides a straightforward way to obtain critical indices, both for the 
bulk and surface parts. At and above the critical dimension $d=6$, it is
found that the percolation cluster contains only a finite number of surface
points in the infinite-size limit. This is in accordance with the
expectation from studies of lattices with negative curvature. It is also
found that the number of surface points, reached by the percolation cluster
in the infinite limit, approaches $2d$ for large dimensions $d$. We also
note that the size dependence in proliferation of percolating
clusters for $d\ge 7$ can be obtained by solely counting surface points
of the midpoint cluster.
\end{abstract}


\maketitle

\section{Introduction}
\label{sec:intro}

Percolation has been extensively studied over the past several decades and
remains as an active field of research. In addition to its
intrinsic scientific value and its role as one of the basic models of critical
phenomena~\cite{chris}, it has contributed to improving our general
understanding of statistical physics on various geometries (see, e.g.,
Ref.~\cite{doro}), as well as developing efficient numerical algorithms
(for a general introduction to percolation, see Ref.~\cite{stauffer}).
A percolation transition can be manifested in
many different ways: a common quantity used in studies of the transition is 
the largest cluster size, but many other quantities also give a clear signature of the transition,
including the cluster size distribution~\cite{leath1,leath2} and
the ratio between the first and second largest cluster sizes~\cite{silva}.
At the same time, percolation is often a question of
connectivity, so one obvious question is then how many connections can be
randomly broken before the system fails to percolate from one side of the
system to the other, or from its middle to the boundary.
From this viewpoint, a percolation phenomenon
requires a surface in order to be meaningful, just as
water has to percolate all the way through the coffee layer
in a coffee percolator. Thus, one might ask how
the percolation transition is reflected on the actual surface.
Specifically, the fraction of surface points belonging to the percolating
cluster, $\mathcal{C}_\infty$, can be written as $P_s \sim
(p-p_c)^{\beta_s}$ for $d$-dimensional systems where $p$ is the occupation
probability and $p_c$ is its critical value. This formulation is parallel to
the bulk criticality, i.e., $P \sim (p-p_c)^{\beta}$ where $P$ is the
fraction of bulk points belonging to $\mathcal{C}_\infty$.
The exponent $\beta_s$ is known to be $4/9$ from conformal
invariance for $d=2$~\cite{vans-vand-roux}, and the mean-field value, valid
for $d\ge 6$, is $\beta_s=3/2$~\cite{carton}. For $d=3$, Monte Carlo
simulations have estimated $\beta_s/\nu = 0.9753(3)$ where $\nu$
describes the correlation length as $\xi \sim (p-p_c)^{-\nu}$~\cite{deng05b}.

The midpoint percolation turns out to be a useful concept
to understand percolation transitions in curved structures~\cite{baek}.
This earlier work focused on lattices with negative Gaussian
curvatures and it was found that the percolation for such lattices contains
two critical thresholds: the first one at which the number of points reached on
the surface from the midpoint becomes finite in the limit of the large
system size, and the second one, where this number becomes a finite fraction
of the surface points.
The reason for the occurrence of two percolation thresholds
appears to be intimately related to the size of the surface: when the
surface-volume ratio is finite, as for the negatively curved lattices,
there appear two thresholds. These two separate thresholds coalesce into a
single one if the surface-volume ratio vanishes. In a $d$-dimensional system
with size $N$,
this ratio is to leading order $N^{-1/d}$. Thus from this viewpoint, a
negatively curved structure with a constant surface-volume ratio is
infinite-dimensional.

Let $b$ be the average number of surface points reached from the midpoint in
a $d$-dimensional hypercubic lattice having a linear size $L$.
The quantity $b$ has a size dependence of the form $b \sim L^{\kappa}$ at
criticality.
The exponent $\kappa$ can be related to the bulk exponent $\beta$ and the
surface exponent $\beta_{s}$ as follows:
\begin{equation}
\kappa = (d-1) - (\beta + \beta_s)/\nu.
\label{eq:kappa}
\end{equation}
This follows since the probability that the percolation cluster at
criticality contains the midpoint, in the large-$L$ limit, is independent
of how many surface points it contains. Thus, the fraction between the number of
surface points and the total number of points of the percolating cluster
becomes $L^{\kappa}/L^{d-1} \sim L^{-\beta/\nu} L^{-\beta_s/\nu}$, where
$L^{-\beta_s/\nu}$ is proportional to $P_s$ and $L^{-\beta/\nu}$ is
proportional to $P$ (see Sec.~\ref{ssec:bulk} below).
Consequently, the exact value $\beta_s = 4/9$ for $d=2$ implies that 
$\kappa = 9/16 = 0.5625$~\footnote{ We
previously conjectured in Ref.~\cite{baek} that $\kappa = 4/7 \approx
0.5714$ for $d=2$ by relating this exponent to the fractal dimension of the
hull~\cite{sapoval}. In the present work, we can distinguish between $4/7$
and $9/16$ and thus verify that the latter is correct.}.
A second case where the exponent $b$ can be obtained analytically is
$d=6$, for which Eq.~(\ref{eq:kappa}) gives $\kappa=0$ with the known values
for the exponents~\cite{carton}. This implies that only a
finite number of points on the surface are reached from the midpoint at
criticality. A third example is
the Cayley tree with coordination number $z$, which is a negatively curved
lattice and hence can be regarded as corresponding to $d=\infty$.
This structure has the lower and upper critical thresholds at
$p_{c1}=1/(z-1)$ and
$p_{c2}=1$, respectively~\cite{baek}, and there exist indefinitely many
percolating clusters between these two thresholds~\cite{benjamini}.
One can calculate the average number of surface
points reached from the midpoint at the lower threshold $p_{c1}$ which gives
$b = z/(z-1)$. Since there is no size dependence, this means that $\kappa =
0$. The implication is that one only reaches a finite number of surface
points from the midpoint for all dimensions $d\ge 6$.

One might also imagine a gradual reduction in the magnitude of the curvature
until the structure becomes an ordinary $d$-dimensional lattice.
Since both of the thresholds coalesce for a regular lattice, $p_c = p_{c1} =
p_{c2}$, we are left with the following possibilities at criticality:
$b$ can be either a positive constant as it is for $p_{c1}$ in case of the
Cayley tree, or an increasing function of $L$ as it is at $p_{c2}$.
In this work, we find that the divergent behavior
of $b$ indeed becomes weaker as $d$ increases, and reaches the
limit of $b=const.$ at the upper critical dimension, $d=6$ in accordance
with the expectation above.
We also show that the surface observable, $b$, provides us with a direct
way of quantifying the percolation transition in general dimensions. In
Sec.~\ref{sec:leath}, we
describe the Leath algorithm used throughout
this work. In Sec.~\ref{sec:res}, we analyze the
numerical results and discuss the critical behavior. Finally,
Sec.~\ref{sec:sum} gives a summary.

\section{Leath algorithm}
\label{sec:leath}
The measurement of $b$ is well suited for the Leath
algorithm~\cite{leath1,leath2}, which is basically a burning algorithm
to create a cluster starting from the midpoint. The collection of burned
sites at each realization will be called a midpoint cluster, and abbreviated as
$\mathcal{C}_m$.
It suffices to count the number of surface points which are contained in
$\mathcal{C}_m$.
This makes the algorithm in the present case easier
to use than in most earlier works since we need no explicit treatment to
exclude the boundary.
Let us consider a $d$-dimensional hypercubic lattice having the length of
each edge as $L=2^n$ with an integer $n$. For convenience, we will impose
the periodic boundary condition in this lattice.
Each point will be assigned a coordinate with $d$
components. Fixing one of the components, one gets a $(d-1)$-dimensional
{\em surface} across the system which serves as our effective boundary layer:
we do not allow the midpoint cluster to grow beyond this layer.
In this way, we effectively get an odd number of lattice sites in each
direction which uniquely defines the midpoint.
 
As for the actual implementation of the burning algorithm, one may choose
between two options: depth-first and breadth-first methods. The former can
be easily made using recursion as follows.
\begin{enumerate}
\item Assign the ``boundary'' state to the points constituting the
boundary layers, and set all the other points as ``unexamined''.
\item Set the midpoint as ``occupied''.
\item For each of the neighboring unexamined points of this occupied one,
\begin{enumerate}
\item with probability $p$, mark the neighbor as occupied and repeat Step
3 with respect to this newly occupied point.
\item With probability $1-p$, mark the neighbor as ``stopped''.
\end{enumerate}
\end{enumerate}
The midpoint cluster consists of the resulting occupied points and its growth
will stop when it is completely surrounded by points marked as either
stopped or boundary.

In spite of the ease of implementation, this method requires
that the program remember the state of every lattice point, which severely
restricts accessible system sizes. For this reason, the other option, i.e.,
the breadth-first method is better suited for the present context. The program is then
only required to remember the outmost shell of $\mathcal{C}_m$. This can be
implemented in the following way.
\begin{enumerate}
\item Prepare a queue and an array, both initialized as empty. The queue
stores actively burning points with its possible directions to proceed,
while the array stores points which cannot be burned again, i.e., either
``stopped'' or currently ``active''.
\item Add the midpoint to the queue, with every direction allowed, and
record the midpoint as well as its state as active in the array.
\item Retrieve an element from the queue. The queue becomes shortened by one
in length on the retrieval. Also delete this active point from the array.
\item For every possible neighbor from the retrieved point,
make a search in the array unless it belongs to the boundary layers.
\begin{enumerate}
\item If it is absent,
\begin{enumerate}
\item with probability $p$, add this to the array as active. Also add
this to the queue with preventing it from propagating back.
\item With probability $1-p$, add this to the array as ``stopped''.
\end{enumerate}
\item If it is found as active, tell the corresponding element in the
queue not to propagate toward this direction.
\item If it is found as stopped, this neighbor is not penetrable. Do nothing.
\end{enumerate}
\item Go to Step 3.
\end{enumerate}
Note that we need not predefine the connection structure since one
can easily compute neighboring coordinates from a given point in regular
structures.
As recommended in earlier works~\cite{lorenz,paul,grass}, we have
employed hashing~\cite{knuth} in Step 4, but not included
the data blocking method~\cite{lorenz} nor the generation of a random number
from each site index~\cite{voll,paul}.
We choose the burning probability $p$ from the previously known
site-percolation thresholds,
and the values used are tabulated in Table.~\ref{table:pc}.
Since $p_c$ decreases with $d$, the actually generated cluster
is much smaller than the effective lattice size in a high
dimension. We also note how efficiently the Leath algorithm
performs with respect to memory usage.
For example, our implementation with 2~200 megabytes memory
can readily simulate a case where
boundaries are away from the midpoint by $31$ lattice spacings in a
six-dimensional hypercube around the known threshold, $p_c \approx
0.109017$~\cite{grass}.
If we loaded every lattice point on memory, the memory requirement should
roughly amount to $2^{36}$ integers, a significant portion of which would
be simply redundant since the burning probability is low.
The depth-first method has been used in
this work only for simulating small system sizes, as well as performance
checks of the breadth-first method.

\section{Results}
\label{sec:res}

\begin{figure}
\includegraphics[width=0.48\textwidth]{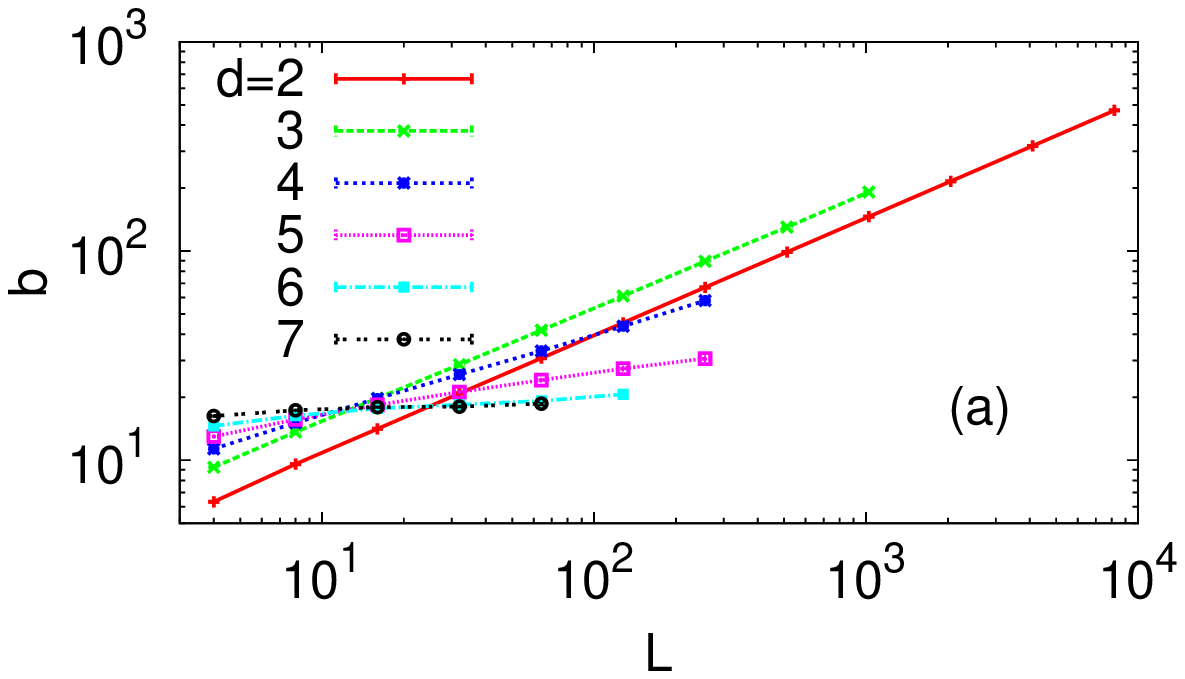}
\includegraphics[width=0.48\textwidth]{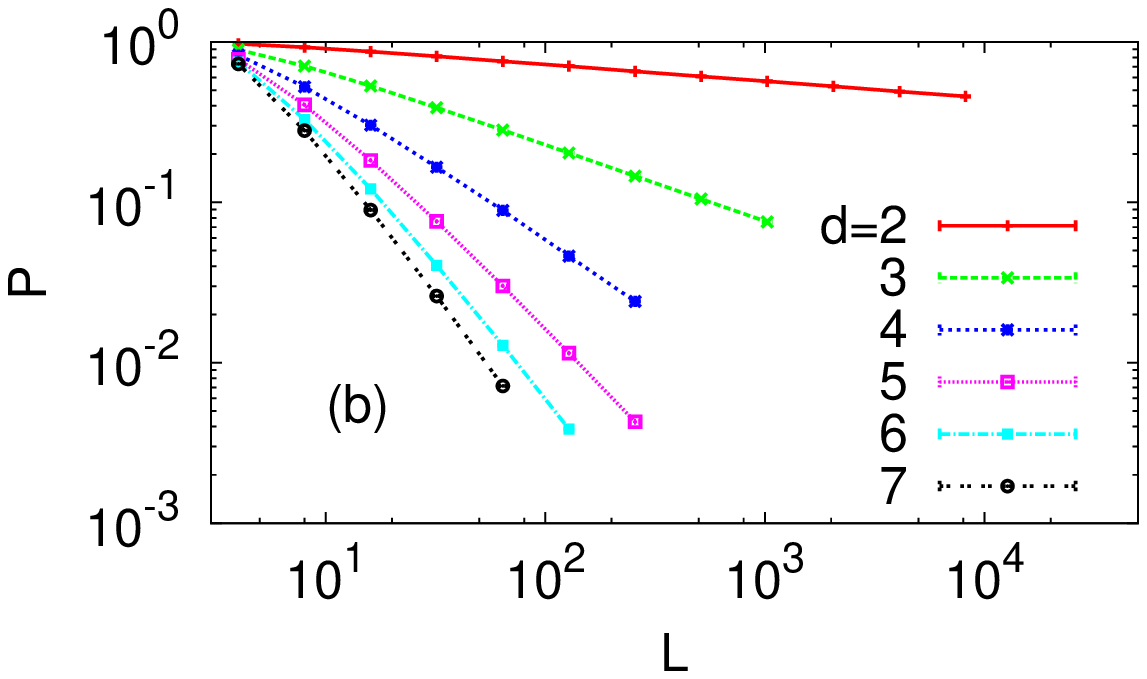}
\includegraphics[width=0.48\textwidth]{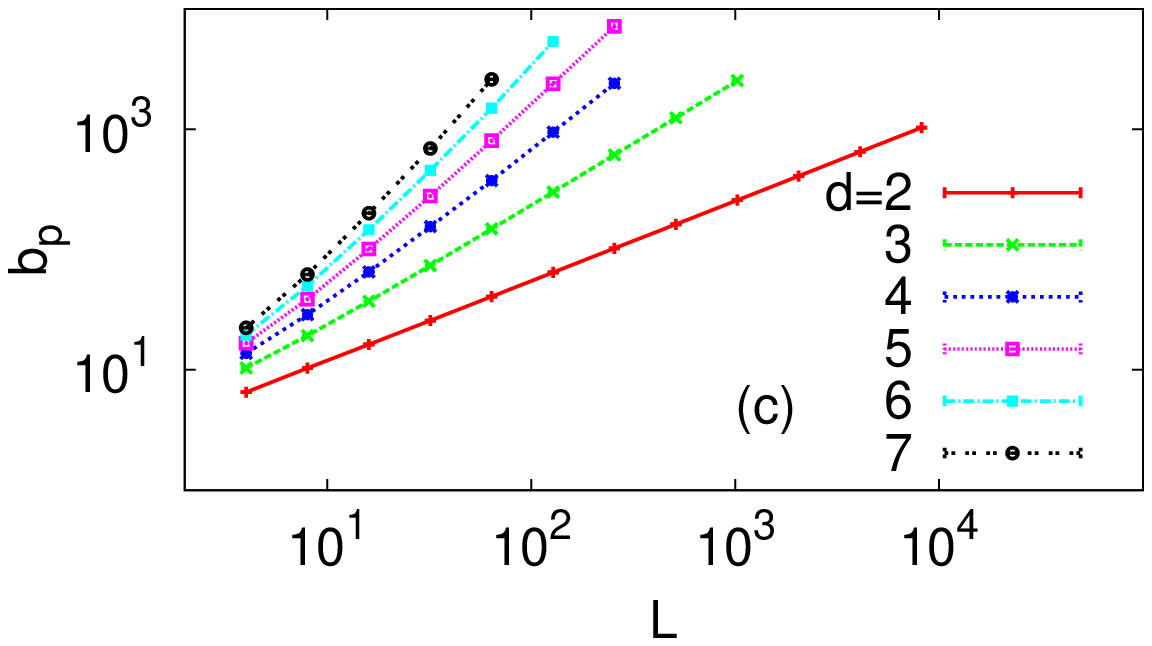}
\caption{(Color online) Data obtained by the Leath algorithm from $d=2$ to $7$.
(a) The average number of surface points, $b$, reached by the midpoint
cluster. (b) The average probability for the midpoint cluster to reach the
boundary, which is identified with $P$, the density of the percolating
cluster. (c) The average number of surface points contained in a percolating
midpoint cluster, $b_p$, for different sizes $L$. Error bars are shown in
all the figures, but usually comparable to the symbol sizes.}
\label{fig:data}
\end{figure}

Using the above algorithms, we have generated more than $10^6$ samples for
each $L$ and $d$ obtainable within our resources. Our numerical results are
presented in Figs.~\ref{fig:data} and \ref{fig:binf} and Table~\ref{table:pc}.
After recording the number of surface points reached by each midpoint cluster,
we carry out the following analysis.

\subsection{Bulk criticality}
\label{ssec:bulk}

Let us consider a lattice at an occupation probability $p$, where one or
more clusters are distributed over the system. Among them, our algorithm
always picks only one cluster, $\mathcal{C}_m$, the midpoint cluster.
When $p$ reaches $p_c$, a percolation cluster $\mathcal{C}_\infty$
will appear among many clusters in the system. It may or may not
contain the midpoint.
In case that the midpoint happens to be contained in $\mathcal{C}_\infty$,
our algorithm shows that the generated cluster, $\mathcal{C}_m$,
actually percolates the system, i.e., reaches the boundary. If
$\mathcal{C}_\infty$ does not contain the midpoint, on the other hand, our
algorithm picks up only a nonpercolating cluster with no surface points.
In short, the frequency of touching the boundary in our algorithm simply
means the probability for a percolating cluster $\mathcal{C}_\infty$ to
contain the midpoint.
Moreover, since the midpoint is one of the lattice points in the bulk, the
probability that it is contained within the percolating cluster
$\mathcal{C}_\infty$ is the same as for any other bulk point. This
probability is given by the density of lattice points in
$\mathcal{C}_\infty$ and consequently has
the size scaling $P \sim L^{d_f -d} = L^{-\beta/\nu}$ where $d_f$ is
the fractal dimension of the percolating cluster. This means that one can
estimate $\beta/\nu$ by
counting how frequently $\mathcal{C}_m$ with a given size $L$
reaches the boundary since when it touches, it does belong to the percolating
cluster with certainty and vice versa. We generally
expect a size scaling of an observable $Q$ at criticality as
\begin{equation}
Q  = A L^{\Upsilon} \left( 1 + B L^{-\Delta} + \cdots \right) + C,
\label{eq:scale}
\end{equation}
where $A$, $B$, and $C$ are $L$-independent constants
and $\Delta > 0$ characterizes the correction to the leading scaling
behavior. In case of $P$, we know that $\Upsilon = -\beta/\nu$.
Furthermore, since $P(L\rightarrow \infty)$ in fact vanishes, we can safely
set $C$ as zero in this case. Therefore, one arrives at
\begin{equation}
P L^{\beta/\nu} \approx A + B' L^{-\Delta}
\label{eq:Ps}
\end{equation}
with $B' \equiv AB$. Since plotting $P L^{\beta/\nu}$ against $L^{-\Delta}$
should give a straight line for correct exponents, we can determine the
exponents from the goodness of fit. For $d=2$, for example, we get
$\beta/\nu = 0.1044(2)$ together with $\Delta = 1.4(2)$, when choosing the
chi-square significance level as $\alpha = 10\%$, and this result agrees
well with the exact value, $\beta/\nu = 5/48 \approx 0.1042$ (see
Ref.~\cite{wu} and references therein). We have employed this method for
lattices of dimension $2\leq d\leq 7$ and the results are presented in
Table~\ref{table:pc} [Fig.~\ref{fig:data}(b)]. Whenever the data fail to
exceed the chi-square critical value (CV) for $\alpha=10\%$, we remove the
smallest system size and then repeat the procedure.
As seen in Table~\ref{table:pc}, the resolution of our data tends to
deteriorate somewhat with increasing dimension (see e.g., the values for
$d=5$). This is presumably caused by the fact that the correction term in
Eq.~(\ref{eq:Ps}) is not enough to absorb all the deviations from the leading
behavior when $L$ is too small. Since the deviations increase with the
dimension, we are restricted to using fewer sizes of $L$ at a larger
dimension.
To some extent, the deviation is also caused by the fact that the number of
available data sets decreases with large $\beta/\nu$, making the statistics
worse. Finally we note that the deviation of $\beta/\nu$ from the predicted
values in case of the upper critical dimension, $d=6$, could possibly also
be attributed to the logarithmic correction~\cite{essam}.

\begin{table*}
\caption{Occupation probabilities used in this work and values of
exponents obtained in this work in comparison to previously known results.
For $d=2$ and $d \ge 6$, the known values of $\beta/\nu$ and $\beta_s/\nu$
are exact. The last column is for a consistency check of our
analysis where the sum should be $d-1$ according to Eq.~(\ref{eq:kappa}).
The discrepancy at $d=7$ signals the proliferation of percolating clusters,
as explained in the text.}
\begin{tabular*}{\hsize}{@{\extracolsep{\fill}}cccccccc}\hline\hline
$d$ & $p_c$ (known) & $\beta/\nu$ (X) & $\beta/\nu$ (known) & $\kappa$ (Y) & $\beta_s/\nu$ (Z) & $\beta_s/\nu$ (known) & X+Y+Z\\\hline

2 & 0.5927460$^{\rm a}$ & 0.1044(2) & 5/48$^{\rm e}$ & 0.563(1) & 
0.33345(10) & 1/3$^{\rm m}$ & 1.001(1)\\
3 & 0.3116081$^{\rm b}$ & 0.478(2) & 0.474(6)$^{\rm f}$ & 0.554(7) &
0.974(2) & 0.970(6)$^{\rm f}$ & 2.01(1)\\
  &                       &          & 0.477(4)$^{\rm g}$ & & & &\\
  &                       &          & 0.4770(2)$^{\rm h}$ & & & &\\
  &                       &          & 0.481(1)$^{\rm i}$ & & &
0.975(4)$^{\rm i}$ &\\
  &                       &          & & & & 0.9754(4)$^{\rm n}$ &\\
  &                       &          & 0.4774(1)$^{\rm j}$ & & &
0.9753(3)$^{\rm j}$ &\\
4 & 0.196889$^{\rm c}$ & 0.945(5) & 0.9528(14)$^{\rm b}$ & 0.410(1) &
1.64(2) & & 3.00(2)\\
  &                      &          & 0.953(7)$^{\rm c}$ & & &\\
5 & 0.1407966$^{\rm d}$ & 1.5(1) & 1.462(16)$^{\rm k}$ & 0.11(7) &
2.408(5) & & 4.0(2)\\
  &                        &  & 1.46(1)$^{\rm c}$ & & &\\
6 & 0.109017$^{\rm d}$ & 1.9(1) & 2$^{\rm l}$ & 0.06(2) &
2.8(3) & 3$^{\rm o}$ & 4.8(4)\\
7 & 0.0889511$^{\rm d}$ & 2.0(1) & 2$^{\rm l}$ & 0.06(8) &
3.08(1) & 3$^{\rm o}$ & 5.1(2)\\ \hline\hline

\end{tabular*}
\label{table:pc}
\begin{tablenotes}
\item $^{\rm a}$ Reference~\cite{ziff}.
\item $^{\rm b}$ Reference~\cite{ball}.
\item $^{\rm c}$ Reference~\cite{paul}.
\item $^{\rm d}$ Reference~\cite{grass}.
\item $^{\rm e}$ Reference~\cite{wu}.
\item $^{\rm f}$ Reference~\cite{gr92}.
\item $^{\rm g}$ Reference~\cite{lorenz}.
\item $^{\rm h}$ Reference~\cite{ball99}.
\item $^{\rm i}$ Reference~\cite{deng}.
\item $^{\rm j}$ Reference~\cite{deng05b}.
\item $^{\rm k}$ Reference~\cite{adler}.
\item $^{\rm l}$ Reference~\cite{harris}.
\item $^{\rm m}$ Reference~\cite{vans-vand-roux}.
\item $^{\rm n}$ Reference~\cite{deng05a}.
\item $^{\rm o}$ Reference~\cite{carton}.
\end{tablenotes}
\end{table*}

\subsection{Surface criticality}

As for the bulk critical exponent $\beta$, we use the surface data in order
to obtain $\kappa$. This time the observable is $b$, the number of surface
points reached by the midpoint cluster $\mathcal{C}_m$
[Fig.~\ref{fig:data}(a)]. As discussed in Sec.~\ref{sec:intro}, the constant
$C$ is expected to be a nonvanishing constant for $d\ge 6$ since it
corresponds to the number of surface points reached in the infinite-size
limit. Thus, according to Eq.~(\ref{eq:scale}), we assume the size scaling
as
\begin{equation}
(b-C) L^{-\kappa} \approx A + B' L^{-\Delta},
\label{eq:bs}
\end{equation}
which is of the same form as Eq.~(\ref{eq:Ps}) apart from
the additional constant, $C$. As a practical data-fitting procedure, we find
the smallest $C$ that gives a sufficiently high CV to pass $\alpha=10\%$.
This procedure yields $\kappa = 0.563(1)$ for $d=2$, which is entirely 
consistent with the exact result $\kappa = 9/16 = 0.5625$.
The result for $2\le d \le 7$ are given in Table~\ref{table:pc}.
One way to obtain $\beta_s/\nu$ is then to use Eq.~(\ref{eq:kappa}) and the
determined values for $\beta/\nu$ and $\kappa$. However, we will here use this
connection as a consistency check as shown in the last column in
Table~\ref{table:pc}.
Instead, we use the alternative method of counting
the average number of surface points reached by the midpoint cluster which
actually do reach the boundary. This means that we only sample over the
cases
when $\mathcal{C}_m$ does percolate [Fig.~\ref{fig:data}(c)]. For
$d=2$, this gives $\beta_s/\nu = 0.33345(10)$, which is close to
the exact value, $1/3$. Table~\ref{table:pc} shows $\beta_s/\nu$ estimated
in this way for $2\le d \le 7$. To our knowledge, the values for
four and five dimensions are reported for the first time in this work.
It is notable that for $d=6$ both $\beta_s/\nu$ and  $\beta/\nu$ deviate
somewhat from the 
theoretical predictions but are still inside the estimated bounds.
Nonetheless, the consistency check $\beta/\nu+\beta_s/\nu+\kappa=d-1$ is
well born out for $2\le d \le 6$, verifying the internal consistency of our
method and analysis (the last column in Table~\ref{table:pc}). We stress
that the three exponents $\beta/\nu$, $\beta_s/\nu$, and $\kappa$ in the
present work are all obtained by just counting the number of surface points
reached by the midpoint percolation cluster at criticality. 

The case of $d=7$ in Table~\ref{table:pc} is of special interest because
$\beta/\nu+\beta_s/\nu+\kappa=5.1(2)< d-1=6$. This means that the relation
$L^{\kappa}/L^{d-1} \sim L^{-\beta/\nu} L^{-\beta_s/\nu}$ breaks down in
this case. The reason is that the percolating cluster at criticality is no
longer unique for $d\ge 7$~\cite{arcan}. This is usually referred to as the
breakdown of the hyperscaling relation~\cite{stauffer}.
Suppose that the number of percolation
cluster scales as $L^x$, then the relation changes to $L^{\kappa}/L^{d-1}
\sim L^{-x} L^{-\beta/\nu} L^{-\beta_s/\nu}$ so that the consistency
relation becomes $\beta/\nu+\beta_s/\nu+\kappa=d-1-x$.
The reason is that the chance that the midpoint belongs to one of the
percolating clusters is still given by $L^{d_f-d}$, but now $L^{-\beta/\nu}$
gives the probability that a lattice point belongs to \emph{any} percolating
cluster. Consequently, the chance for the midpoint to be contained in one of
percolating clusters is
$L^{-x} L^{-\beta/\nu}$ from which $\beta/\nu+\beta_s/\nu+\kappa=d-1-x$
follows. The number of percolating clusters is expected to scale as $L^{d-6}$
for $d\geq 6$~\cite{arcan}. The consistency relation for $d>6$ then
becomes $\beta/\nu+\beta_s/\nu+\kappa=5$. As seen in Table~\ref{table:pc},
this relation is born out by our results for $d=7$. We also observe that,
provided that the mean-field values hold for the  critical indices $\beta/\nu$
and $\beta_s/\nu$ for $d\geq 6$, and in addition the critical index
$\kappa=0$ for $d\geq 6$ (as implied by our results), then it follows from
the consistency relation that $x=d-6$.
Conversely, if we take the growth of the percolation
clusters, $L^{d-6}$, for granted, then we obtain $\kappa=0$ from the
consistency relation. This also implies that the fractal
dimension of a single percolating cluster becomes
$d_f=d-x-\beta/\nu=4$ for $d\geq 6$~\cite{arcan}.
It is interesting to note that the
exponent $x$, describing how the number of percolating clusters grows
with a lattice size at criticality for $d\geq 6$ can be obtained
by just counting the number of surface points reached by the midpoint
cluster.

\begin{figure}
\includegraphics[width=0.48\textwidth]{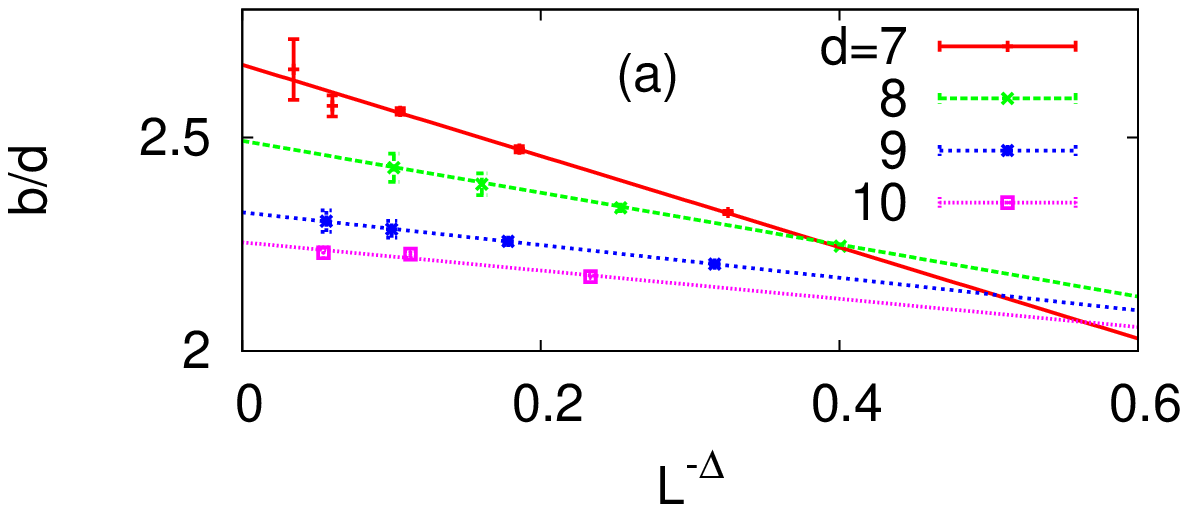}
\includegraphics[width=0.48\textwidth]{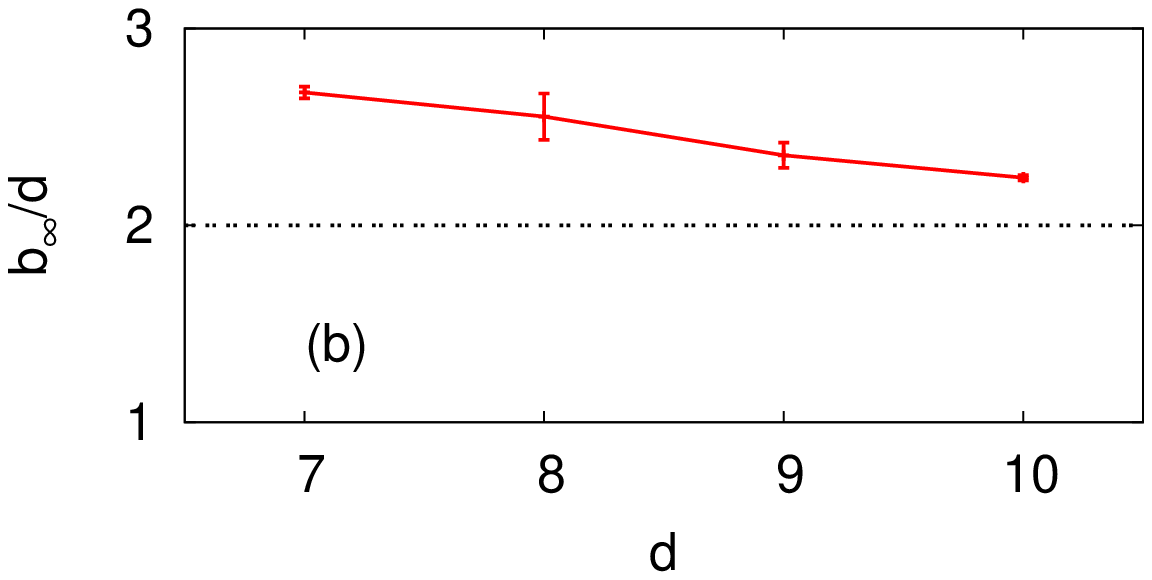}
\includegraphics[width=0.48\textwidth]{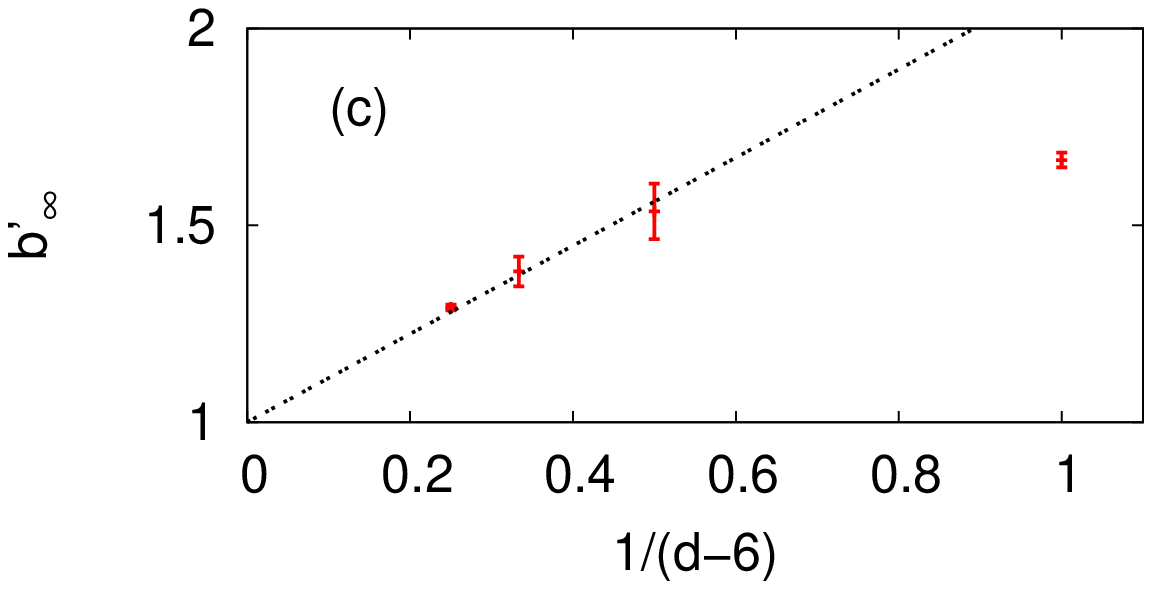}
\caption{(Color online) (a) Measurements of $b/d$ for dimensions from $d=7$
to $10$, where $b$ is fitted by Eq.~(\ref{eq:bs2}). The correction exponents
are estimated as $\Delta (d=7) = 0.5(1)$, $\Delta (d=8) = 0.8(2)$, $\Delta
(d=9) = 0.7(4)$, and $\Delta (d=10) \gtrsim 1$.
(b) The limiting number of surface points, $b(L \rightarrow \infty)$,
divided by $d$. (c) The number of surface points reached, when
percolation along the surface is prohibited, as a function of
$1/(d-6)$. Linear extrapolation suggests that $b'_\infty(d \rightarrow \infty)
\approx 1$.}
\label{fig:binf}
\end{figure}

Next we assume that the inferred result $\kappa=0$ for $d \ge 6$ is correct.
Then Eq.~(\ref{eq:bs}) reduces to
\begin{equation}
b \approx b_\infty + B' L^{-\Delta},
\label{eq:bs2}
\end{equation}
where $b_\infty \equiv A+C$. This says that $b (L
\rightarrow \infty)$ approaches a constant denoted as $b_\infty$. We have
estimated this constant up to $d=10$, using the critical probabilities
reported in Ref.~\cite{grass} [Fig.~\ref{fig:binf}(a)].
The correction exponent $\Delta$ shows a tendency to
increase as $d$ grows so that the convergence to $b_\infty$ becomes more
rapid. Figure~\ref{fig:binf}(b) suggests that the limiting value of $b_\infty$
for large $d$ could possibly be $2d$. In order to investigate this further,
we note that $b'(L)=p_cb(L-1)$ corresponds to the number of points reached
on the surface \emph{provided} that percolation along the surface is
prohibited. In Fig.~\ref{fig:binf}(c), we have plotted $b'_\infty \equiv
b'(L\rightarrow \infty)$ against $1/(d-6)$, where
we assume that $d-6$ is a fundamental parameter in the problem. Linear
extrapolation suggests that the limiting value for $d\rightarrow \infty$ is
close to $b'_{\infty}=1$. Since $p_c$ approaches $1/(2d-1)$ for large $d$,
this result also implies that $b_\infty \approx 2d$, even though the
precision is not sufficient to make any firm conclusion.
Note that $2d$ is just
the number of faces in a $d$-dimensional hypercube, so that one can say
that $\mathcal{C}_m$ reaches every face of the surface in one point on the
average. However, any individual realization of the midpoint percolation
cluster will of course reach a variety of points on each face and some not
at all.

In a previous study, it was shown that the cluster grows on the
\emph{average} by one site per
step for large enough dimensions, suggesting a connection to the
self-avoiding random walk (SAW)~\cite{grass}: A cluster that grows with
\emph{precisely} one site per step traces out a SAW. From this perspective,
it is interesting to note that our quantity $b'_\infty$ corresponds to the
average number of points that a SAW walker starting from the midpoint
reaches on the
surface. Since a SAW walker always has a finite chance of getting stuck in
any dimension $d<\infty$, this means that $b'_\infty<1$ for SAW, whereas we
have found $b'_\infty>1$ for the midpoint percolation cluster at
criticality. However, the chance of getting stuck vanishes for the
SAW walker in the limit of $d\rightarrow \infty$, which means that
$b'_\infty=1$ in this limit. This agrees with our corresponding
result for percolation, as well as with the result in Ref.~\cite{grass},
suggesting some similar feature between an unhindered SAW and the midpoint
cluster at criticality in the limit of large $d$. However, the cluster
created by SAW and the midpoint percolation cluster at criticality
have quite different structures in all dimensions including the limit of
$d\rightarrow \infty$ since the SAW cluster has fractal dimension $d_f =2$
for high $d$ whereas that of the midpoint percolation cluster is $d_f=4$
as verified in the present paper. One may note that if one only takes the
backbone of the percolating cluster, the backbone indeed has a fractal
dimension $d_f^{(b)}=2$ for $d>6$~\cite{stauffer}, which strengthens the
similarity.

\section{Summary}
\label{sec:sum}

We studied the percolation transitions in dimensions from $d=2$ to
$10$ by the Leath algorithm. In particular, it was shown that one can
obtain various critical properties by just counting the number of points
reached on the surface from the midpoint.
To this end, we checked that our midpoint percolation yielded consistent
results with known ones, and estimated exponents characterizing the surface
criticality as well as the bulk one.
We found that the divergent behavior of $b$ becomes
weaker in higher dimensions as anticipated from the Cayley tree, so that it
scales as $L^{9/16}$ for $d=2$ but converges to a constant in the
infinite-size limit for $d \ge 6$.
We also confirmed that the percolation cluster ceases to be unique for
$d\geq 6$.
In addition, our results suggests that the number of surface points reached
approaches the value of $2d$ in the limit of large $d$. Provided that
percolation along the surface excluded, this corresponds to the simple result
that precisely one surface point is reached at criticality.

\acknowledgments
S.K.B. and P.M. acknowledge the support from the Swedish Research Council
with the Grant No. 621-2002-4135, and
B.J.K. was supported by the Korea Research Foundation Grant funded
by the Korean  Government(MEST) with the Grant No. KRF-2008-005-J00703.
This research was conducted using the resources of High Performance
Computing Center North (HPC2N).


\end{document}